# Unstable Vicinal Crystal Growth from Cellular Automata

A. Krasteva[1, a)], H. Popova[2, b)], F. Krzyżewski[3, c)], M. Załuska-Kotur[3,4 d)] and V. Tonchev[2, e)]

[1]*Institute of Electronics, BAS, 72 Tzarigradsko chaussee blvd, 1784 Sofia, Bulgaria*
[2]*Institute of Physical Chemistry, BAS, Acad. G. Bonchev Str., block 11, 1113 Sofia, Bulgaria*
[3]*Institute of Physics, Polish Academy of Sciences, al. Lotników 32/46, 02-668 Warszawa, Polska, Warsaw, Poland.*
[4]*Faculty of Mathematics and Natural Sciences, Card. S. Wyszynski University, ul. Dewajtis 5, 01-815 Warsaw, Poland*

[a)]Corresponding author: anna0kr0stz@gmail.com
[b)] karleva@ipc.bas.bg
[c)] fkrzy@ifpan.edu.pl
[d)] zalum@ifpan.edu.pl
[e)] tonchev@ipc.bas.bg

**Abstract.** In order to study the unstable step motion on vicinal crystal surfaces we devise vicinal Cellular Automata. Each cell from the colony has value equal to its height in the vicinal, initially the steps are regularly distributed. Another array keeps the adatoms, initially distributed randomly over the surface. The growth rule defines that each adatom at right nearest neighbor position to a (multi-) step attaches to it. The update of whole colony is performed at once and then time increases. This execution of the growth rule is followed by compensation of the consumed particles and by diffusional update(s) of the adatom population. Two principal sources of instability are employed – biased diffusion and infinite inverse Ehrlich-Schwoebel barrier (iiSE). Since these factors are not opposed by step-step repulsion the formation of multi-steps is observed but in general the step bunches preserve a finite width. We monitor the developing surface patterns and quantify the observations by scaling laws with focus on the eventual transition from diffusion-limited to kinetics-limited phenomenon. The time-scaling exponent of the bunch size $N$ is 1/2 for the case of biased diffusion and 1/3 for the case of iiSE. Additional distinction is possible based on the time-scaling exponents of the sizes of multi-step $N_{\text{multi}}$, these are 0.36÷0.4 (for biased diffusion) and 1/4 (iiSE).

## INTRODUCTION

Surface patterning is still subject of significant interest with steady shift from fundamental to more application-oriented contributions [1, 2], the main reason behind being the quest for nano-templates to be used in the controllable bottom-up synthesis of nano-structures. These developments challenge the modelling community and rather sophisticated models were developed based on the Monte Carlo methods [3, 4]. These are still restricted in modelling systems with sizes that permit good statistics at affordable price. Thus new models are to be developed that are simpler and more flexible in reproducing the general features of the studied systems.

Cellular Automata (CA) are expected to bring a "new kind of science" [5]. We restrict ourselves from commenting the Wolfram's viewpoint but still we use CA as a simple yet powerful tool to study situations for which the other approaches are insufficient. In order to be able to access some methodological problems in step bunching (SB) studies, such as the universality classes [6] in SB and the classification [7] of SB, we devise new CA to model the unstable motion of steps on vicinal crystal surfaces. Sources of instability that are employed are biased diffusion and iiSE - infinite barrier for attachment of the adatoms to the steps from the lower terrace.

Our strategic goal is to develop models in higher dimensions that would account for all factors influencing the growing morphologies to help the morphologies fine-tuning.

# THE MODEL

In order to achieve atomic scale resolution with retained possibility for fast calculations on large systems we devise our vicinal Cellular Automata. Each cell from the 1D colony is given a value equal to its height in the vicinal stairway that descends from the left to the right. In the beginning, the steps are regularly distributed at distance $l_0$. Another 1D array of the same size *NAS* contains the adatoms. In the beginning they are randomly distributed over the surface with concentration *c*. The growth rule defines that each time there is an adatom at the right nearest neighboring site to a (multi-) step it attaches to the step. Then the lowest layer of the (multi-)step advances one position to the right done by increasing the height of the vicinal at the position of the adatom by 1. The adatom is deleted from the adatoms array. The growth updates using the growth rule are performed at once – the update of each cell from the vicinal is kept in a mirror array while every cell is thus updated, then the whole cell population is updated at instance using this "mirror" array and then the time is increased by 1. Each execution of the automaton rule is complemented by compensation of the adatom concentration to *c* and the adatom population is then subjected to diffusional update(s), their number is denoted by $n_{DS}$. In any diffusional update a total number of adatoms equal to the size of the array *NAS* is chosen at random, then tried to jump left or right with some probabilities, usually their sum being 1, except in the case of iiSE. These updates do not lead to a time change. Thus, with increasing $n_{DS}$ is realized a transition from diffusion-limited growth to a kinetics-limited one – diffusing adatoms make many hops before being eventually captured by the growing surface. The diffusion is influenced by one of two principal sources of instability - bias or iiSE. Note the iiSE turns the model into one-sided. The bias is realized by defining that the hop probability to the right as $(0.5 + b)$ while to the left it is $(0.5 - b)$. The iiSE is realized through inhibition of the diffusional hops to the left when the adatom is right next nearest neighbor to a (multi-)step and inhibition of the diffusional hops to the right when the adatom is right nearest neighbor to a (multi-)step. These destabilizing factors are not opposed by step-step repulsions leading to the formation of multi-steps.

The model permits fast calculations on systems with large sizes thus achieving the regime of intermediate asymptotics where a reliable statistics is collected for the monitored properties. In order to monitor the developing surface patterns we adopt a modification of an established monitoring protocol [8], it is the formation of multi-steps that determines the need of this modification. Important criterion build into it defines when two neighboring steps belong to the same bunch - when the distance between them is less than $l_0$. Four quantities describing the developing instability are monitored – the terrace width *Tw*, defined as the distance between two sequential bunches, the size (number of bunch distances) *N* and width *W* of the bunches and the size of the multi-steps $N_{multi}$[8]. In this short article we continue the program aimed at distinguishing the diffusion-limited [9] from kinetics-limited [10] regimes of SB and describe the instabilities in both cases (biased diffusion and iiSE) by proper scaling laws.

# NUMERICAL RESULTS

Here are presented first qualitative data on the surface profiles and step trajectories and then, in the second part, quantitative data on time-scaling behavior of the representative quantities – bunch size *N*, bunch width *W*, terrace width *Tw* and size of the multi-steps $N_{multi}$. These data are collected in order to account for the possible transition from diffusion-limited to kinetics-limited growth regime with changing the number of diffusional steps $n_{DS}$.

## Qualitative Results – Step Trajectories and Surface Profiles

In Figure 1 are shown qualitative data on the surface profiles and step trajectories for the case of iiSE. These results are collected for two values of the numbers of diffusional updates per growth one, $n_{DS} = 1$ (Fig. 1a, b) and $n_{DS} = 80$ (Fig. 1c, d). The formation and the time evolution of step bunches for $n_{DS} = 1$ and $n_{DS} = 80$ are shown in Fig. 1a and Fig. 1c, respectively. It is clearly seen that the bunches consist of both multi- and mono-steps. The bunches obtained with $n_{DS} = 80$ have larger tails thus they have generally larger widths and this observation is confirmed further by the quantitative results. In Figs. 1b, d are presented surface profiles for well-developed instability.

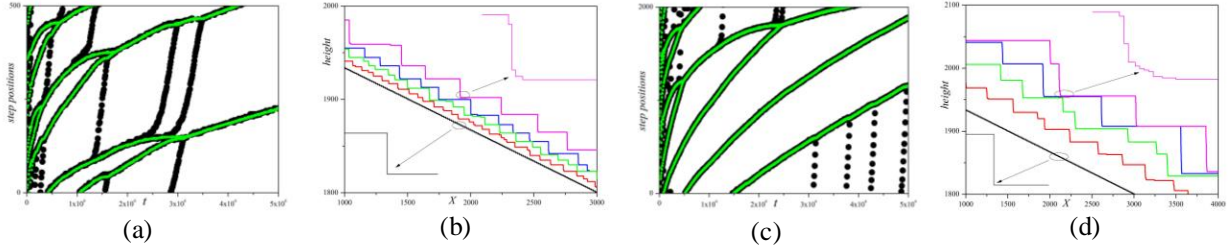

**FIGURE 1**. Trajectories multi- and mono-steps denoted by green and black dots (a, c) and surface profiles (b, d) for two values of the diffusional steps per growth one: $n_{DS} = 1$ (a, b) and $n_{DS} = 80$ (c, d). The initial vicinal distance $l_0 = 15$ and the concentration $c = 0.2$ are the same for the four plots.

## Quantitative Results for the Surface Time Evolution

In Figure 2 are presented data on time evolution of the bunch size $N$, bunch width $W$, terrace width $Tw$ and size of the multi-steps $N_{multi}$ for the case of biased diffusion. The time scaling exponent of the average bunch size $N$ is always 1/2. This universal behavior is demonstrated further in the inset where $N/n_{DS}^{1/2}$ is plotted versus time (see Fig. 2a), the data collapse on a single line. The bunch widths (Fig. 2b) remain finite throughout the simulation with a slight increase. The time scaling exponent of the size of multi-steps is smaller than the one of of the bunch size, showing that multi-steps increase slower than the bunches in which the multi-steps are, in fact, embedded. It ranges from 0.36 (Fig. 2b) to 0.4 (not shown). Note that the data for $N_{multi}$ are not rescaled with the values of $n_{DS}$. Time scaling of the terrace width $Tw$ also shows a universal behavior with changing the number of diffusional steps per growth one and shows the same time-scaling exponent of 1/2 as the one of the bunch size $N$.

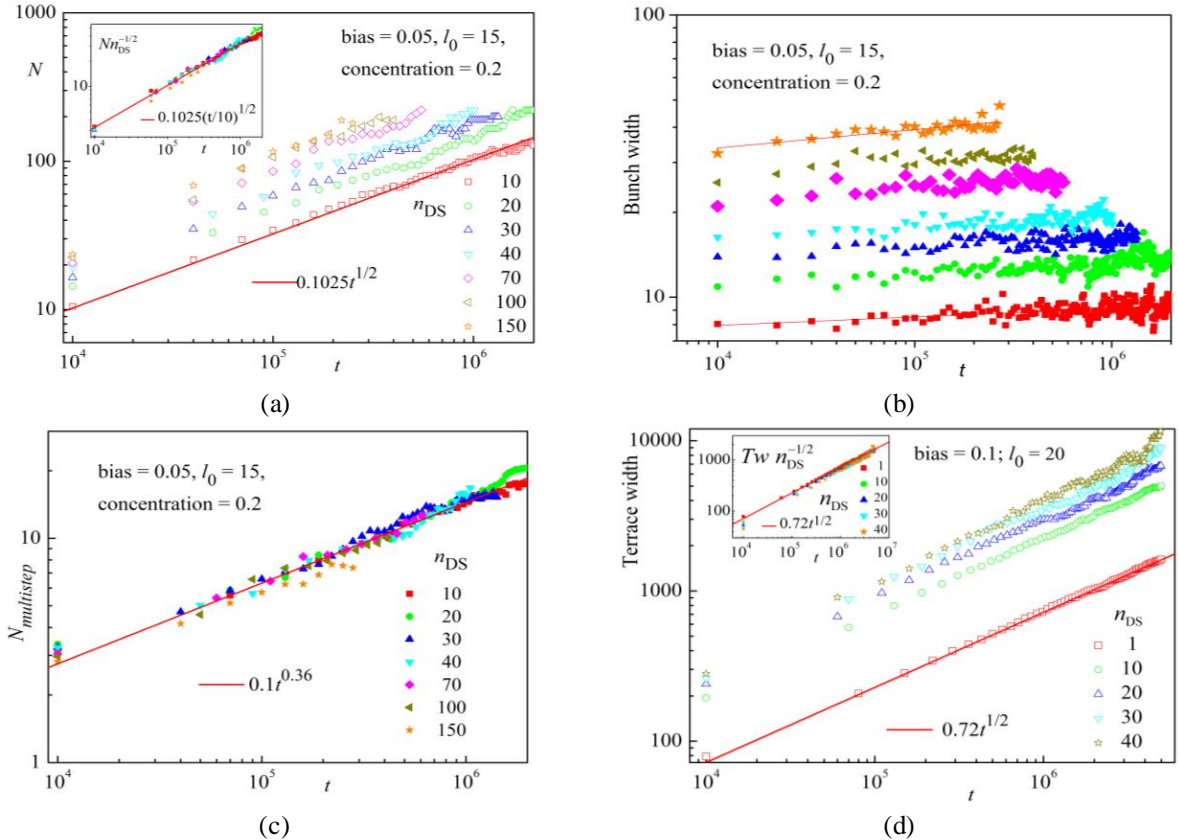

**FIGURE 2**. Time-scaling of (a) the average bunch size $N$, (b) the bunch widths $W$, (c) multistep size $N_{multi}$ for values of $n_{DS}$ ranging from 10 to 150 and $bias = 0.05$, $l_0 = 15$, $c = 0.2$ and (d) time-scaling of the terrace width $Tw$ for values of $n_{DS}$ ranging from 1 to 40 for $bias = 0.1$, $l_0 = 20$, $c = 0.2$.

In Figure 3 are presented data for the time evolution of the bunch size $N$ and of the size of multi-steps $N_{multi}$ for the case of iiSE. The time-scaling exponent of the bunch size $N$ is 1/3 different from the case of biased diffusion where it is 1/2 and remains universal for $n_{DS}$ ranging from 1 to 100 (Fig. 3a). Time-scaling exponent of $N_{multi}$ also differs from the one for bunch size and equals 0.25, providing a ground of additional differentiation between the two cases. Note the data are not rescaled with $n_{DS}$ (Fig. 3b). Time scaling of the $Tw$ is not shown here since it has the same exponent as the bunch size and scales in the same way with $n_{DS}$, the behavior of the bunch width is similar to the one for the case of bias.

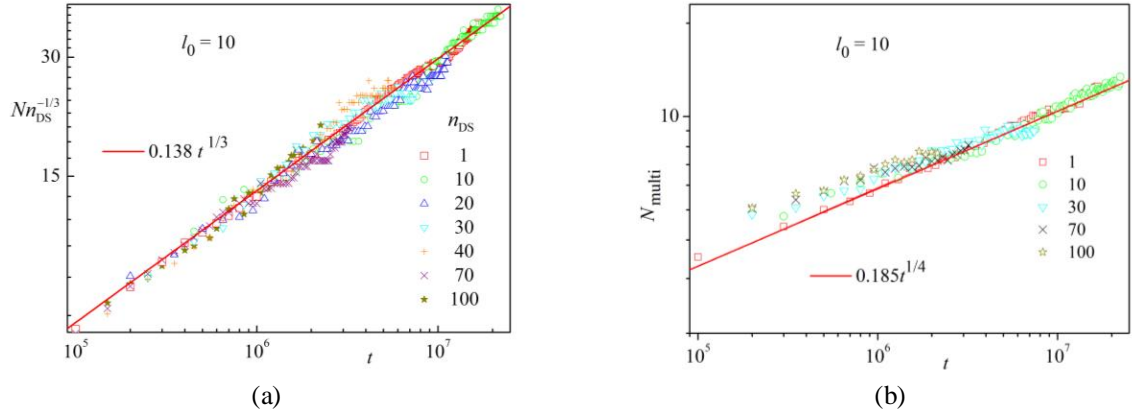

(a) (b)

**FIGURE 3**. Time-scaling of the average bunch size $N$ (a) and of average multistep size $N_{multi}$ (b) for values of $n_{DS}$ ranging from 1 to 100 for $l_0$ =10, c = 0.2 in case of iiSE.

## CONCLUSION

In order to model surface patterning in unstable vicinal growth we build a new model based on the Cellular Automata and employ it to study two cases of instability with principal sources being biased diffusion and iiSE. We focused here on the possible transition from diffusion-limited to kinetics-limited regime of growth by increasing the number of diffusional steps that follow the growth one. We found that the scaling laws describing quantitatively the instability differ for both cases but within each case account for both regimes, thus remaining universal.

## ACKNOWLEDGMENTS

This work is supported by Bulgarian NSF, grant No. T02-8/121214, and a bilateral BAS-PAS project. F.K. acknowledges NCN of Poland, grant No.2013/11/D/ST3/02700, and a fellowship under Erasmus+.